\documentclass[prb,twocolumn,superscriptaddress,floatfix,showpacs,showkeys,preprintnumbers,amssymb,amsmath,a4]{revtex4}
\usepackage[latin1]{inputenc}
\usepackage{graphicx}% Include figure files
\usepackage{dcolumn}% Align table columns on decimal point
\usepackage{bm}% bold math
\usepackage[mathscr]{eucal}
\usepackage{epsfig}
\usepackage{rotating}

\begin{document}

\newcommand{\beq}{\begin{equation}}
\newcommand{\eeq}{\end{equation}}
\newcommand{\ket}{\rangle}
\newcommand{\bra}{\langle}
\newcommand{\A}{\mathbf{A}}
\preprint{ }
\title{Tunable coupling scheme for flux qubits at the optimal point}

\author{Antti O. Niskanen}
\email{niskanen@frl.cl.nec.co.jp}
\affiliation{CREST-JST, Kawaguchi, Saitama 332-0012,Japan}
\affiliation{VTT Technical Research Centre of Finland, Sensors, PO BOX 1000, 02044 VTT, Finland}
\author{Yasunobu Nakamura}
\affiliation{CREST-JST, Kawaguchi, Saitama 332-0012,Japan}
\affiliation{NEC Fundamental Research Laboratories, Tsukuba, Ibaraki 305-8501, Japan}
\affiliation{The Institute of Physical and Chemical Research (RIKEN), Wako, Saitama 351-0198, Japan}

\author{Jaw-Shen Tsai}
\affiliation{CREST-JST, Kawaguchi, Saitama 332-0012,Japan}
\affiliation{NEC Fundamental Research Laboratories, Tsukuba, Ibaraki 305-8501, Japan}
\affiliation{The Institute of Physical and Chemical Research (RIKEN), Wako, Saitama 351-0198, Japan}

\date{\today}

\begin{abstract}
We discuss a practical design for tunably coupling a pair of flux qubits
via the quantum inductance of a third high-frequency qubit.
The design is particularly well suited for realizing 
a recently proposed microwave-induced parametric coupling scheme.
This is attractive because the qubits can always remain at their optimal points.
Furthermore, we will show that the resulting coupling also has an optimal point where it is insensitive to low-frequency flux noise.
This is an important feature for the coherence of coupled qubits. The presented scheme is an experimentally realistic way of carrying out
two-qubit gates and should be easily extended to multiqubit systems.
\end{abstract}


\maketitle

\section{Introduction}

Superconducting qubits\cite{schon} have received a lot of attention during the last few years. 
The observed coherence times have been improved from the initial ns\cite{nec1} range to the $\mu$s\cite{saclay,delft,delft2,wallraff,martinis} range
or so with the invention of so-called optimal bias points. That is, it is possible to set the biases of a qubit such that
the energy difference of the utilized computational states do not depend on bias value to, say, first order. This makes
the system at hand insensitive to harmful low-frequency noise. 
It seems that if an optimal bias point exists then the qubits should always be biased there\cite{delft2,ithier}.

However, isolated qubits are not usable for quantum computing. Some type of controllable coupling between individual qubits is desirable. 
Typically tunable coupling schemes\cite{schon,averin,you,filippov,cosmelli,berkeley,napoli,newjournal,liu} 
require moving away from the optimal bias point. 
This is because the ``natural'' coupling --- inductive for flux qubits and capacitive for charge qubits --- is off-diagonal 
(${\sigma_x\otimes\sigma_x}$) at the optimal
point and affects the qubit dynamics only in second order provided that the qubits are detuned. 
Therefore, constant moderate coupling can often be neglected at the optimal point.
Recently, however, two promising schemes for 
coupling superconducting qubits (flux or charge) at the optimal point have emerged. In the so-called
FLICFORQ coupling scheme\cite{flicforq} two ${\sigma_x\otimes\sigma_x}$-coupled detuned qubits can be made to interact by applying 
resonant microwaves on the qubits at a power such that the sum of the Rabi frequencies of the individual qubits 
matches the detuning between the qubits. This causes a kind of a second resonance in the interaction picture, which can be used to realize
a universal two-qubit gate. However, the scheme is challenging to realize currently with flux qubits since achieving small enough detunings
in the fabrication is hard and they easily exceed achievable Rabi frequencies. 
Nevertheless, because of the above reasons, Bertet {\it et.~al.}\cite{bertet} suggested an alternative parametric coupling scenario which relies
on the modulation of the coupling energy itself at the sum or the difference frequency of two detuned qubits.
Liu {\it et.~al.}\cite{liu} suggested earlier a related microwave coupling scheme which does not unfortunately work at the optimal point.
The gained benefit in the scheme of Bertet {\it et.~al.} compared to FLICFORQ 
is that the method is in principle applicable to qubits with a larger detuning and it
may be possible to carry out gates faster.
This scheme of course has an analogy for 
charge qubits too, but was suggested primarily for flux qubits. We also concentrate on flux qubits in this paper. 
The problem then is what kind of a physical 
coupling element to use to achieve the parametric control of the coupling energy. In the original paper\cite{bertet} a method utilizing a 
current biased DC-SQUID\cite{berkeley}
was considered to exemplify the scheme. This requires the modulation of the bias current at a high frequency. 
Moreover, the DC current bias of the SQUID must
be near switching current and the flux near half flux quantum. This condition however contradicts with the optimal current bias 
condition of individual qubits\cite{delft2} since the bias current noise linearly couples to the qubits
at such bias condition.

We suggest using an extra high-frequency qubit for the parametric AC-modulated coupling. 
The used circuit can be otherwise identical to the primary qubits,
but the splitting should be larger so that the third qubit is always in the ground state. 
The functioning of this coupling can be viewed also in terms of the so-called quantum inductance. 
In fact the present scheme is very similar to using a Cooper pair box\cite{averin}
as a coupling element or using an RF-SQUID\cite{newjournal} both of which schemes were suggested as 
ordinary DC coupling.
The use of a third qubit has many nice features. 
First of all, it should be easy to fabricate this kind of circuitry since the technology is compatible. It is easy to ensure that
the third qubit indeed has higher frequency than the other two.
Secondly, the control can be all magnetic and frequency multiplexed (assuming different splittings) with a minimal number of microwave lines. 
Also, it turns out that the effective coupling at DC can be set to zero if desired. This however
may not be necessary because of the ${\sigma_x\otimes\sigma_x}$-nature of the coupling. What is more important
is that a kind of an optimal point exists for the effective coupling energy. This may be crucial since
at this point the effective coupling energy is insensitive to low-frequency flux noise, and therefore two-qubit oscillations
can be expected to be long lasting. The time to generate a universal entangling gate similar to CNOT is on the order of 10-100 ns with
experimentally realistic parameters.

The paper is divided as follows. In Sec.~\ref{sec:ham} we first derive an approximate Hamiltonian starting from the three-qubit 
Hamiltonian. The meaning of the different terms will be discussed. Then in Sec.~\ref{sec:acscheme} we discuss the implementation of the 
coupling scheme in the context of the present system. In Sec.~\ref{sec:sim} we present results of numerical simulations.
There the validity of the effective Hamiltonian is tested in simulation by comparing the dynamics of the truncated system 
with that of the full system. Section~\ref{sec:disc} is dedicated to discussion.

%% \onlinecite
\section{System and the Hamiltonian}\label{sec:ham}
\begin{figure}
\begin{picture}(130,160)
\put(-55,0){\includegraphics[width=0.45\textwidth]{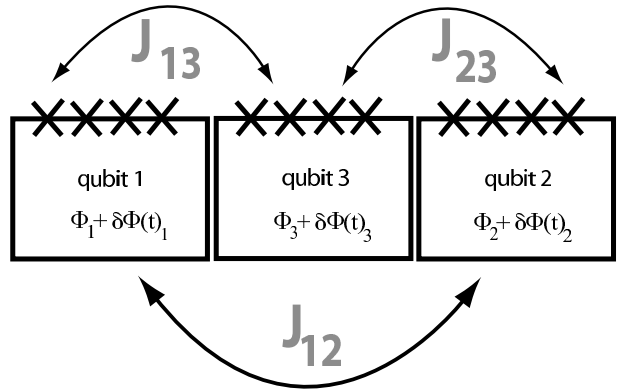}}
\end{picture}
\caption{\label{fg:schematic}
Schematic illustration of three anti-ferromagnetically coupled flux qubits.
}
\end{figure}

We consider a system of three inductively coupled flux qubits as illustrated in Fig.~\ref{fg:schematic}.
The qubits consist of, say, three\cite{delft} or four\cite{delft2} ideally identical junctions
except for one junction that has an area smaller by a factor $\alpha$.
The large junctions are characterized by a charging energy $E_{\rm C}=e^2/2C$ where $C$ is the junction capacitance and by a Josephson energy
$E_{\rm J}=\hbar I_{\rm C}/2e$ where $I_{\rm C}$ is the critical current of the junction.
Ideally, the smaller junction has a charging energy $E_{\rm C}/\alpha$ and a Josephson energy $\alpha E_{\rm J}$. 
Typically, $E_{\rm J}/E_{\rm C} \sim 10^1 -10^2$. For three-junction qubits $\alpha$ is around 0.75  and for four-junction qubits around 0.5.
These choices ensure that a two-well potential forms and furthermore, a two-level (qubit) approximation of the system is valid.
We neglect the effect of loop inductances in the single qubit energies, which should be valid as long as Josephson inductance dominates. For a rigorous
derivation of the coupled-qubit Hamiltonian, see Ref.~[\onlinecite{alec}].

The three-qubit Hamiltonian reads in the persistent current basis
\beq\label{eq:fullham}
H_{3q}=H_{3DC}+H_{3MW}.
\eeq
The DC part of the Hamiltonian reads
\begin{equation}
H_{3DC}=-\frac{1}{2}\sum_{j=1}^{3}\left(\Delta_j \sigma_x^j+\epsilon_j \sigma_z^j \right)
-\frac{1}{2}\sum_{k\neq l}J_{kl}\sigma_z^k\sigma_z^l,
\end{equation}
where $\epsilon_j=2I_{{\rm p} j}(\Phi_j-\Phi_0/2)$, $I_{{\rm p} j}$ is the maximum persistent current of the qubit $j$, $\Phi_j$ is the
flux threading the loop of the $j^{\rm th}$ qubit and $\Phi_0=h/2e$ is the flux quantum. The approximation is valid near the optimal point
for the qubits, i.e., $\Phi_j\approx \Phi_0/2$. 
The tunneling amplitudes $\Delta_j$
are defined by the Josephson energies and capacitances of the qubit junctions. These splittings depend in particular exponentially
on $\alpha$.
Moreover, the couplings are defined by the mutual inductances $M_{kl}$, i.e., 
$J_{kl}=J_{lk}=M_{kl}I_{{\rm p}k} I_{{\rm p}l}$. The couplings can be either ferromagnetic ($M_{kl}>0$) or anti-ferromagnetic ($M_{kl}<0$)
depending on the geometry. 
The microwave controlled Hamiltonian simply reads
\beq
H_{3MW}=-\frac{1}{2}\sum_{j=1}^{3}\delta\epsilon_j(t) \sigma_z^j.
\eeq
Here $\delta\epsilon_j(t)=2I_{{\rm p} j}\delta\Phi_j(t)$,
and $\delta\Phi_j(t)$ is the AC component of flux threading the loop of the qubit $j$.

Let us assume that $\omega_1,\omega_2\ll \omega_3$, where 
$\omega_j=\sqrt{\Delta_j^2+\epsilon_j^2}$.
Applying the unitary transformation 
\beq\label{eq:U}
\tilde{U}=\frac{1}{\sqrt{2}}\left(\sqrt{1+\frac{\epsilon_3(t)}{\omega_3(t)}}I_3+i\sqrt{1-\frac{\epsilon_3(t)}{\omega_3(t)}}\sigma_y^3\right)
\eeq
yields ($\tilde{H}_{3q}=\tilde{U}H_{3q}\tilde{U}^\dagger$)
\begin{align}\label{eq:adiaham}
\tilde{H}_{3q}=-\frac{1}{2}\sum_{j=1}^{2}\left(\Delta_j \sigma_x^j+\epsilon_j \sigma_z^j \right)-\frac{1}{2}\omega_3(t)\sigma_z^3
-J_{12}\sigma_z^1\sigma_z^2\nonumber \\
-(J_{13}\sigma_z^1+J_{23}\sigma_z^2)\left(\frac{\epsilon_3(t)}{\omega_3(t)}\sigma_z^3-\frac{\Delta_3}{\omega_3(t)}\sigma_x^3\right)
-\frac{1}{2}\sum_{j=1}^{2}\delta\epsilon_j(t) \sigma_z^j.
\end{align}
We denote the identity operation on qubit $j$ by $I_j$.
Here we have included the temporal dependence of the bias of the third qubit in the transformation, i.e., we are transforming to the adiabatic 
basis of the third qubit.  That is $\epsilon_3(t)=\epsilon_3+\delta\epsilon_3(t)$
and $\omega_3(t)=\sqrt{\epsilon_3(t)^2+\Delta_3^2}$. We will use an adiabatic approximation for the third qubit.

By looking at Eq.~(\ref{eq:adiaham}) we see that in the lowest order adiabatic approximation for the third qubit,
i.e., when $\sigma_z^3\to {\bra\sigma_z^3\ket}=1$ and $\sigma_x^3\to{\bra\sigma_x^3\ket}=0$, the flux biases of the qubits 1 and 2 will be shifted.
The $\sigma_x^3$-component will have no effect on the dynamics of the two qubits to lowest order.
But going to higher order in $J_{13}$ and $J_{23}$ will yield an effective coupling term.
In order to eliminate the third high-frequency qubit we use a trick known as the Schrieffer-Wolff transformation (see e.g. Ref.~[\onlinecite{martti}]).
We look for a transformation $\exp(-S)$ such that the antihermitian operator $S=-S^\dagger$ is first order in $J_{kl}$ 
such that it eliminates the  $\sigma_x^3$-component in first order. That is, we require
\beq\label{eq:wolffcond}
[S,H_0]=H_1+O(J_{kl}^2)
\eeq
where 
\begin{align}
H_0&=-\frac{1}{2}\sum_{j=1}^{2}\left(\Delta_j \sigma_x^j+\epsilon_j \sigma_z^j \right)-\frac{1}{2}\omega_3(t)\sigma_z^3-J_{12}\sigma_z^1\sigma_z^2
\end{align}
and
\begin{align}
H_1=(J_{13}\sigma_z^1+J_{23}\sigma_z^2)\frac{\Delta_3}{\omega_3(t)}\sigma_x^3.
\end{align}
If Eq.~(\ref{eq:wolffcond}) is satisfied we get up to second order
\begin{align}
&\exp(-S)(H_0+H_1)\exp(S)\nonumber \\
&=H_0+H_1+[H_0+H_1,S]+\frac{1}{2}[[H_0,S],S]+O(J_{kl}^3)\nonumber \\
&=H_0+(H_1-[H_0,S])+\frac{1}{2}[H_1,S]+O(J_{kl}^3)
\end{align}
such that the term $H_1$ is eliminated to first order since $(H_1-[H_0,S])$ is second order.
The total Hamiltonian has also the terms
\beq
H_2=-(J_{13}\sigma_z^1+J_{23}\sigma_z^2)\frac{\epsilon_3(t)}{\omega_3(t)}\sigma_z^3
\eeq
and
\beq
H_3=-\frac{1}{2}\sum_{j=1}^{2}\delta\epsilon_j(t) \sigma_z^j.
\eeq

It turns out that Eq.~(\ref{eq:wolffcond}) is solved by choosing
\beq
S=i\alpha\sigma_y^3+i\beta\sigma_x^3,
\eeq
where $\alpha$ and $\beta$ contain only operators operating on the qubits 1 and 2.
A straightforward calculation gives up to first order
\begin{align}
\alpha&=\frac{J_{13}\Delta_3}{\omega_3(t)^2-\Delta_1^2-\frac{\Delta_1^2\epsilon_1^2}{\omega_3(t)^2-\epsilon_1^2}}\sigma_z^1\nonumber \\
&+\frac{J_{23}\Delta_3}{\omega_3(t)^2-\Delta_2^2-\frac{\Delta_2^2\epsilon_2^2}{\omega_3(t)^2-\epsilon_2^2}}\sigma_z^2\nonumber \\
&-\frac{J_{13}\Delta_3\Delta_1\epsilon_1}{(\omega_3(t)^2-\Delta_1^2)(\omega_3(t)^2-\epsilon_1^2)-\Delta_1^2\epsilon_1^2}\sigma_x^1\nonumber \\
&-\frac{J_{23}\Delta_3\Delta_2\epsilon_2}{(\omega_3(t)^2-\Delta_2^2)(\omega_3(t)^2-\epsilon_2^2)-\Delta_2^2\epsilon_2^2}\sigma_x^2.
\end{align}
and
\begin{align}
\beta&=\frac{J_{13}\Delta_1\Delta_3\omega_3}{(\omega_3(t)^2-\Delta_1^2)(\omega_3(t)^2-\epsilon_1^2)-\Delta_1^2\epsilon_1^2}\sigma_y^1\nonumber \\
&+\frac{J_{23}\Delta_2\Delta_3\omega_3}{(\omega_3(t)^2-\Delta_2^2)(\omega_3(t)^2-\epsilon_2^2)-\Delta_2^2\epsilon_2^2}\sigma_y^2.
%&+\frac{J_{13}J_{12}\Delta_3\Delta_1\epsilon_1}{\omega_3((\omega_3(t)^2-\Delta_1^2)(\omega_3(t)^2-\epsilon_1^2)-\Delta_1^2\epsilon_1^2)}
%\sigma_y^1\sigma_z^2 \nonumber \\
%&+\frac{J_{23}J_{12}\Delta_3\Delta_2\epsilon_2}{\omega_3((\omega_3(t)^2-\Delta_2^2)(\omega_3(t)^2-\epsilon_2^2)-\Delta_2^2\epsilon_2^2)}
%\sigma_y^2\sigma_z^1.
\end{align}
Now the effective Hamiltonian reads 
\begin{align}
&\exp(-S)(H_0+H_1+H_2+H_3)\exp(S)\nonumber \\
&\approx H_0+\frac{1}{2}[H_1,S]+H_2+H_3.
\end{align}
The second order terms $[H_2,S]$ and $(H_1-[H_0,S])$ were dropped because they are off-diagonal in the qubit 3 operators.
Since the single-qubit MW-term $H_3$ is anyway weak, we will not consider its transformation. 
We also neglect $dS/dt$ since it vanishes when the adiabatic approximation is made. 
We need to calculate 
\begin{align}
\frac{1}{2}[H_1,S]&=\frac{i}{2}[\frac{\Delta_3}{\omega_3(t)}(J_{13}\sigma_z^1+J_{23}\sigma_z^2),\beta] \nonumber \\
&-\frac{1}{2}\{\frac{\Delta_3}{\omega_3(t)}(J_{13}\sigma_z^1+J_{23}\sigma_z^2),\alpha \} \sigma_z^3,
\end{align}
where $\{\cdot\,,\cdot\}$ stands for the anticommutator. We get
\begin{align}
&\frac{i}{2}[\frac{\Delta_3}{\omega_3(t)}(J_{13}\sigma_z^1+J_{23}\sigma_z^2),\beta]=\nonumber \\
&\frac{J_{13}^2\Delta_1\Delta_3^2}{\omega_3(t)^2(\omega_3(t)^2-\Delta_1^2-\epsilon_1^2)}\sigma_x^1\nonumber \\
&+\frac{J_{23}^2\Delta_2\Delta_3^2}{\omega_3(t)^2(\omega_3(t)^2-\Delta_2^2-\epsilon_2^2)}\sigma_x^2
\end{align}
and (up to constant when the approximation $\sigma_z^3\approx 1$ is made)
\begin{widetext}
\begin{align}
\frac{1}{2}\{\frac{\Delta_3}{\omega_3(t)}(J_{13}\sigma_z^1+J_{23}\sigma_z^2),\alpha \} \sigma_z^3&\approx
\frac{J_{23}J_{13}\Delta_3^2}{\omega_3(t)^3}\left(\frac{\omega_3(t)^2-\epsilon_2^2}{\omega_3(t)^2-\Delta_2^2-\epsilon_2^2}
+\frac{\omega_3(t)^2-\epsilon_1^2}{\omega_3(t)^2-\Delta_1^2-\epsilon_1^2}
\right)\sigma_z^1\sigma_z^2\nonumber \\
&-\frac{J_{23}J_{13}\Delta_3^2}{\omega_3(t)^3}\left(\frac{\Delta_1\epsilon_1}{\omega_3(t)^2-\Delta_1^2-\epsilon_1^2}\sigma_x^1\sigma_z^2
+\frac{\Delta_2\epsilon_2}{\omega_3(t)^2-\Delta_2^2-\epsilon_2^2}\sigma_z^1\sigma_x^2
\right).\nonumber \\
\end{align}
\end{widetext}

We may safely neglect the $\sigma_x^1\sigma_z^2$ and $\sigma_z^1\sigma_x^2$ terms if the third qubit frequency $\omega_3(t)$  
is significantly higher than $\epsilon_j$ and $\Delta_j$ for $j=1,2$.
The effective Hamiltonian for two qubits can thus be compactly rewritten as
\begin{equation}\label{eq:H2}
H_{2q}=-\frac{1}{2}\sum_{j=1}^{2}\left(\tilde{\Delta}_j \sigma_x^j+\tilde{\epsilon}_j \sigma_z^j \right)
-\tilde{J}_{12}(t)\sigma_z^1\sigma_z^2-\frac{1}{2}\sum_{j=1}^{2}\delta\epsilon_j(t) \sigma_z^j.
\end{equation}
with
\beq\label{eq:eps}
\tilde{\epsilon}_j=\epsilon_j+\frac{2\epsilon_3(t)J_{j3}}{\omega_3(t)},
\eeq
\beq\label{eq:delta}
\tilde{\Delta}_j=\Delta_j-
\frac{2(J_{j3}\Delta_3)^2\Delta_j}{\omega_3(t)^2(\omega_3(t)^2-\Delta_j^2-\epsilon_j^2)}
\eeq
and
\begin{align}\label{eq:J}
&\tilde{J}_{12}(t)=J_{12}\nonumber \\
&+\frac{J_{23}J_{13}\Delta_3^2}{\omega_3(t)^3}\left(\frac{\omega_3(t)^2-\epsilon_2^2}{\omega_3(t)^2-\Delta_2^2-\epsilon_2^2}
+\frac{\omega_3(t)^2-\epsilon_1^2}{\omega_3(t)^2-\Delta_1^2-\epsilon_1^2}
\right).
\end{align}
Let us discuss the physical meaning of the different terms in the approximate Hamiltonian in Eq.~(\ref{eq:H2}). 
Recalling that $J_{j3}=M_{j3}I_{{\rm p}j}I_{\rm p3}$ and 
since the ground state expectation value of current for the third qubit is in the absence of the coupling
\beq
\bra 0_3|I_3|0_3 \ket=-\frac{1}{2}\frac{d\omega_3}{d\Phi}=-\frac{I_{p3}\epsilon_3}{\omega_3},
\eeq
we see immediately that the correction in Eq.~(\ref{eq:eps}) $2\epsilon_3(t)J_{j3}/\omega_3(t)=-2M_{j3}I_{{\rm p}j} \bra 0_3|I_3|0_3 \ket $ can be 
interpreted as a shift of the qubit $j$ bias due to the circulating current of the third qubit. 
From now on we consider the modified ``sweet spot'' 
$\tilde{\epsilon}_j=0$. Note that this cannot be perfectly achived at DC
because the time average of $\tilde{\epsilon}_j=0$ depends on the modulation amplitude of the third qubit microwave.
However, the deviation from the ideal case is only second order in $\delta \epsilon_3(t)/\omega_3$, where $\omega_3$ is the DC value
of the third qubit frequency and $\delta\epsilon_3(t)$ is the microwave modulation of the third qubit bias.
Other than that, we neglect the time dependence of $\tilde{\epsilon}_j$ since it can be absorbed in the microwave
Hamiltonian and results in a weak crosscoupling.
Since we are considering a resonant control scheme such omissions are well justified.
The interpretation of the renormalization of $\Delta_j$ is more complex. 

\begin{figure}
\begin{picture}(130,190)
\put(-65,0){\includegraphics[width=0.53\textwidth]{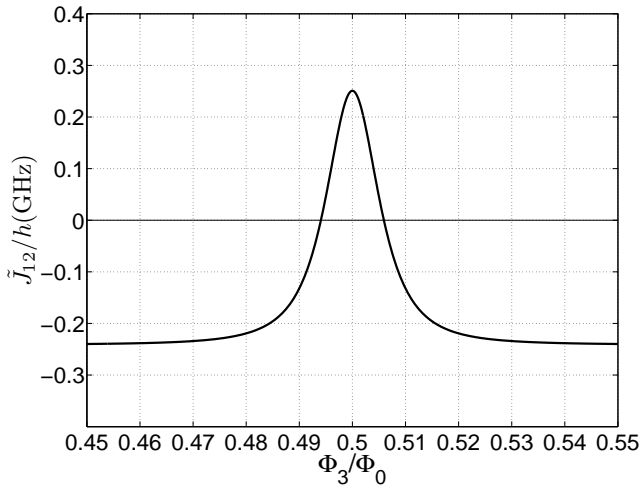}}
\end{picture}
\caption{\label{fg:J12}
The dependence of the effective coupling term on flux when $I_{{\rm p}j}=0.4 \mu$A, $\Delta_3/h=10$ GHz
$\epsilon_2/h=\epsilon_1/h=1.72$ GHz,
$\Delta_1/h=3.1$ GHz, $\Delta_2/h=4.3$ GHz, , $M_{13}=M_{23}=-6$ pH and $M_{12}=-1$ pH.
(For these choices $\tilde{\epsilon}_1=\tilde{\epsilon}_2=0$ if $\epsilon_3/h=7.375$ GHz.) }
\end{figure}

The effective coupling term in Eq.~(\ref{eq:J}) can most easily be understood as coupling through the quantum inductance of the third qubit. 
By quantum inductance we mean the inverse of the coefficient by which the circulating current of the 
auxiliary qubit responds to a change in its bias flux, i.e.
\beq
\frac{1}{L_Q(t)}=\frac{d\bra 0_3|I_3|0_3 \ket}{d\Phi}=-\frac{2I_{{\rm p}3}^2\Delta_3^2}{\omega_3(t)^3}.
\eeq
In the limit of large $\omega_3$ we therefore obtain
\begin{align}
&\tilde{J}_{12}(t)\approx J_{12}+\frac{2J_{23}J_{13}\Delta_3^2}{\omega_3^3}\nonumber \\
&=\left(M_{12}-\frac{M_{23}M_{13}}{L_Q(t)}\right)I_{{\rm p}1}I_{{\rm p}2}.
\end{align}
From this form it is clear that the effective mutual inductance between the qubits is affected by the quantum inductance of the auxiliary qubit; qubit 3
generates a shielding current in response to changes in the flux of qubit 1 (2) which in turn changes the flux through qubit 2 (1).
In order to cancel the effective DC coupling the following conditions must be satisfied since $L_Q<0$: If $M_{12}<0$, then 
$M_{13}$ and $M_{23}$ must have the same sign but if $M_{12}>0$ then $M_{13}$ and $M_{23}$ must have a different sign.
Naturally the magnitudes need to be suitable, too. 
Figure~\ref{fg:J12} illustrates the coupling term as a function of flux with realistic experimental parameters.
Note that two points exist where $g_0=0$.

To conclude the present Section we rewrite the Hamiltonian conveniently at the optimal point $\tilde{\epsilon}_j=0$, via
rotating it by ${\exp(i\pi/4(\sigma_y^1+\sigma_y^2))}$,
as
\begin{equation}\label{eq:Hopt}
H_{2q}^{\rm opt}=-\frac{1}{2}\sum_{j=1}^{2}\left(\tilde{\Delta}_j\sigma_z^j-\delta\epsilon_j(t) \sigma_x^j\right)
-\tilde{J}_{12}(t)\sigma_x^1\sigma_x^2.
\end{equation}
Because at the optimal point observed single qubit coherence times are much superior to those measured elsewhere,
we will focus our attention there.

\section{Implementation of the parametric coupling scheme}\label{sec:acscheme}

As suggested in Ref.~[\onlinecite{bertet}], the form of the Hamiltonian in Eq.~(\ref{eq:Hopt}) is ideal for the realization
of a coupling scheme in which the coupling constant $J_{12}(t)$ is modulated sinusoidally at the angular frequency 
$\omega_\pm=(\tilde{\Delta}_2\pm\tilde{\Delta}_1)/\hbar$. The essence of the scheme is seen easily by considering a general modulation of the form
\beq
\tilde{J}_{12}(t)=g_0+g_+(t)\cos(\omega_+t)+g_-(t)\cos(\omega_-t).
\eeq
Here
\beq
g_\pm\approx \frac{d \tilde{J}_{12}}{d\epsilon_3}\delta\epsilon_{3\pm}
\eeq
where $\delta\epsilon_{3\pm}$ is the amplitude of the modulation of $\epsilon_3(t)$ at either the sum or the difference frequency
and $g_0$ is the DC component of the coupling $\tilde{J}_{12}(t)$ of Eq.~(\ref{eq:J}).
For single-qubit operations we use Rabi oscillations driven by a resonant microwave
\beq
\delta\epsilon_j(t)=2\Omega_j(t)\cos(\tilde{\Delta}_jt/\hbar+\phi_j(t)).
\eeq
In our setup all the temporal dependence of the Hamiltonian is assumed to arise from the time-dependent flux. 
The rotating wave approximation, which is also valid if crosscouplings are taken into account, results
in a rotating frame Hamiltonian of the form
\begin{align}\label{eq:Hrot}
&H_{2q}^{\rm rot}=\frac{1}{2}\sum_{j=1}^{2}\Omega_j(t)\left(\cos\phi_j(t)\sigma_x^j-\sin\phi_j(t)\sigma_y^j\right) \nonumber \\
&-\frac{g_+(t)}{4}(\sigma_x^1\sigma_x^2-\sigma_y^1\sigma_y^2)-\frac{g_-(t)}{4}(\sigma_x^1\sigma_x^2+\sigma_y^1\sigma_y^2).
\end{align}
This can be taken as our logical Hamiltonian. The approximation is valid when $g_0,g_\pm < \omega_\pm$ and $\Omega_j<\Delta_j$.
We must include the possibility of phase shift to get arbitrary one qubit gates, hence the temporal dependence of the microwave phase.
All the control is carried out by a combination of phase and amplitude modulation.
We have set the phase of $\omega_\pm$ to zero for simplicity. 

The Hamiltonian in Eq.~(\ref{eq:Hrot}) is clearly universal. All the single qubit operations can be obtained easily e.g. by rotating three times
around two different axis. From the point of view of two-qubit operations Eq.~(\ref{eq:Hrot})
is extremely nice because it can in principle be used to realize directly the so-called B-gate\cite{bgate}, 
which is known to be the best all-purpose two-qubit gate there is. 
Namely two applications of the B-gate suffice to realize any two-qubit gate.
Essentially, the B-gate is up to some local one-qubit operations given by 
\beq
U_B\sim\exp\left(i\frac{\pi}{4}\left(\sigma_x^1\sigma_x^2+\frac{1}{2}\sigma_y^1\sigma_y^2\right)\right).
\eeq
To realize this with a MW pulse of the length $\Delta t$ one just sets $\Omega_j=0$, $g_+ \Delta t/h=1/8$
and  $g_- \Delta t/h=3/8$.

If the modulation of the coupling is not possible at both the sum and the difference frequencies, say, due to a too small
difference in $\omega_+$ and $\omega_3$, we may still do a universal quantum gate when $g_+=0$. 
The easiest way to achieve this is to do a gate essentially equal (i.e. locally equivalent) to the so-called
double-CNOT, or DCNOT. The nontrivial two qubit part of this gate can be written e.g. as 
\beq\label{eq:DCNOT}
U_{\rm DCNOT}\sim\exp\left(i\frac{\pi}{4}\left(\sigma_x^1\sigma_x^2+\sigma_y^1\sigma_y^2\right)\right).
\eeq
Choosing $\Omega_j=0$, $g_+=0$ and  $ g_-\Delta t/h=1/2$ realizes the gate.
Three applications of $U_{\rm DCNOT}$ with some one-qubit gates suffice to realize any two-qubit gate\cite{optimal}. The gate is thus just as good as 
the better known CNOT.

In order to seriously evaluate the viability of the present scheme, one needs to consider the matter of decoherence.
We are operating in the optimal point of the individual qubits so that there is more hope in this respect than in some
non-optimal point scenario. However, in the presence of coupling, the situation is more complicated.
As has been previously shown\cite{bertet}, if $g_0\ll \omega_-$, the coupling circuit does not add seriously to decoherence,
meaning the single-qubit transition frequencies are stable.
Especially if $g_0=0$, the individual qubits are truly
first order insensitive to fluctuations in all the fluxes $\Phi_j$. 
If $g_0$ is however too large compared to $\omega_-$ then the benefits of the optimal point may unfortunately be lost.

\begin{figure}
\begin{picture}(130,190)
\put(-65,0){\includegraphics[width=0.53\textwidth]{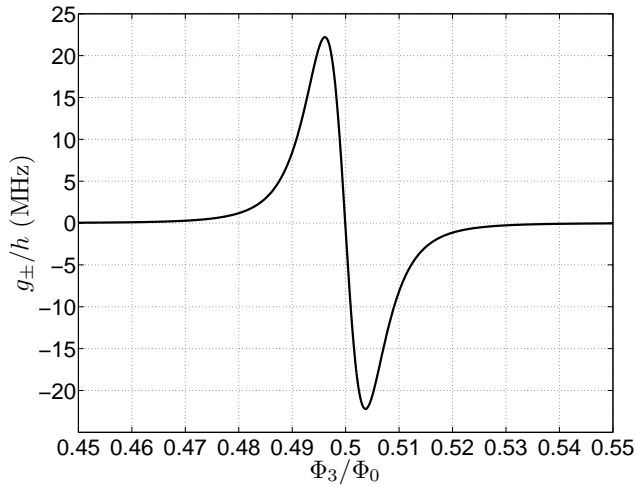}}
\end{picture}
\caption{\label{fg:stab}
The effective AC coupling constant $g_\pm$ as a function of DC flux when the flux amplitude through the third qubit
has the realistic value $\delta\Phi_3=2\times 10^{-4}\Phi_0$.
}
\end{figure}

It is also important to consider the phase coherence of the $|01\ket$ to $|10\ket$
(or $|00\ket$ to $|11\ket$) oscillation. It turns out that our scheme is nice in this sense. Namely, we would like the frequency of oscillation
to be stable against low-frequency noise, i.e. $dg_\pm/d\Phi_3=0$. 
There indeed exist points symmetrically around $\Phi_3=\Phi_0/2$ at which this is satisfied, 
i.e., approximately at $\epsilon_3=\pm\Delta_3/2$. 
Figure~\ref{fg:stab}
illustrates the form of $g_\pm$ with the same parameters as in Fig.~\ref{fg:J12}.
However, there is no reason for these special points to be ones in which $g_0=0$.
For perfect first order insensitivity to 1/f flux noise, it is necessary to design the circuitry so that the conditions  $g_0=0$ and
$dg_\pm/d\Phi_3=0$ are fulfilled simultaneously. Currently the technology may not be there yet though, since in particular $\Delta_3$ is hard to control
in fabrication. It should be however possible to satisfy $g_0\ll \omega_-$ well enough and set still  $dg_\pm/d\Phi_3=0$.
One possible solution is to fabricate qubits with comparably high $\Delta_j$.

Let us compare the proposed scheme with the earlier scheme of coupling through a DC-SQUID.
In our scheme the added improvement
is the existence of the additional stability in the form of $dg_\pm/d\Phi_3=0$. 
Also, the bias conditions are more favorable for the individual qubits.
Although the effect of the DC-SQUID on the single qubit terms
analogous to Eqs.~(\ref{eq:eps}-\ref{eq:delta}) was not included in 
the consideration of coupling through a DC-SQUID in Refs.[\onlinecite{bertet,berkeley}],
the SQUID will couple directly to the individual qubits.
At high bias current and close to $\Phi_0/2$ the bias current noise always couples in first order to $\sigma_x$. 
This is known to especially cause a rapid decrease of the relaxation 
time $T_1$\cite{delft2} and thus also of the dephasing time.
Because of the expected significant decline in $T_1$ we argue that the use of a third high-frequency qubit may be a more viable option
for realizing the promising parametric coupling scheme.

\section{Simulation}\label{sec:sim}
In order to confirm the validity of the approximations made, we present some simulation results. We can conveniently simulate the dynamics
of the system both with the full three-qubit Hamiltonian and with the effective Hamiltonian in Eq.~(\ref{eq:Hopt}).
The form of the full Hamiltonian that we use is the Hamiltonian in Eq.~(\ref{eq:fullham}) rotated by the time-independent version of 
Eq.~(\ref{eq:U}) so that the Hamiltonian is exact but in a more convenient basis. Also the qubits 1 and 2 are rotated. That is,
\begin{align}
&H_{\rm fullsim}=
-\frac{1}{2}\sum_{j=1}^{2}\left(\Delta_j \sigma_z^j-\epsilon_j \sigma_x^j \right)-\frac{1}{2}\omega_3\sigma_z^3\nonumber \\
&-J_{12}\sigma_x^1\sigma_x^2
+(J_{13}\sigma_x^1+J_{23}\sigma_x^2)\left(\frac{\epsilon_3}{\omega_3}\sigma_z^3-\frac{\Delta_3}{\omega_3}\sigma_x^3\right) \nonumber \\
&+\frac{1}{2}\sum_{j=1}^{2}\delta\epsilon_j(t)\sigma_x^j
-\frac{1}{2}\left(\frac{\epsilon_3}{\omega_3}\sigma_z^3-\frac{\Delta_3}{\omega_3}\sigma_x^3\right)\delta\epsilon_3(t).
\end{align}
No rotating wave approximation has been used for the simulation, but rather we solve the full Schr\"{o}dinger equation.
The results are obtained using Floquet states\cite{book} since the problem is time-periodic for the duration of the two-qubit pulse.
Below, the initial state of the qubit 3 is always assumed to be $|0_3\ket$ in the basis of the above Hamiltonian. 
The reduced density matrix $\rho_{1,2}$ is obtained by tracing over the third qubit.
We concentrate here on the use of the difference frequency since this should be experimentally easier to achieve.

\begin{figure}
\begin{picture}(130,200)
\put(-75,0){\includegraphics[width=0.55\textwidth]{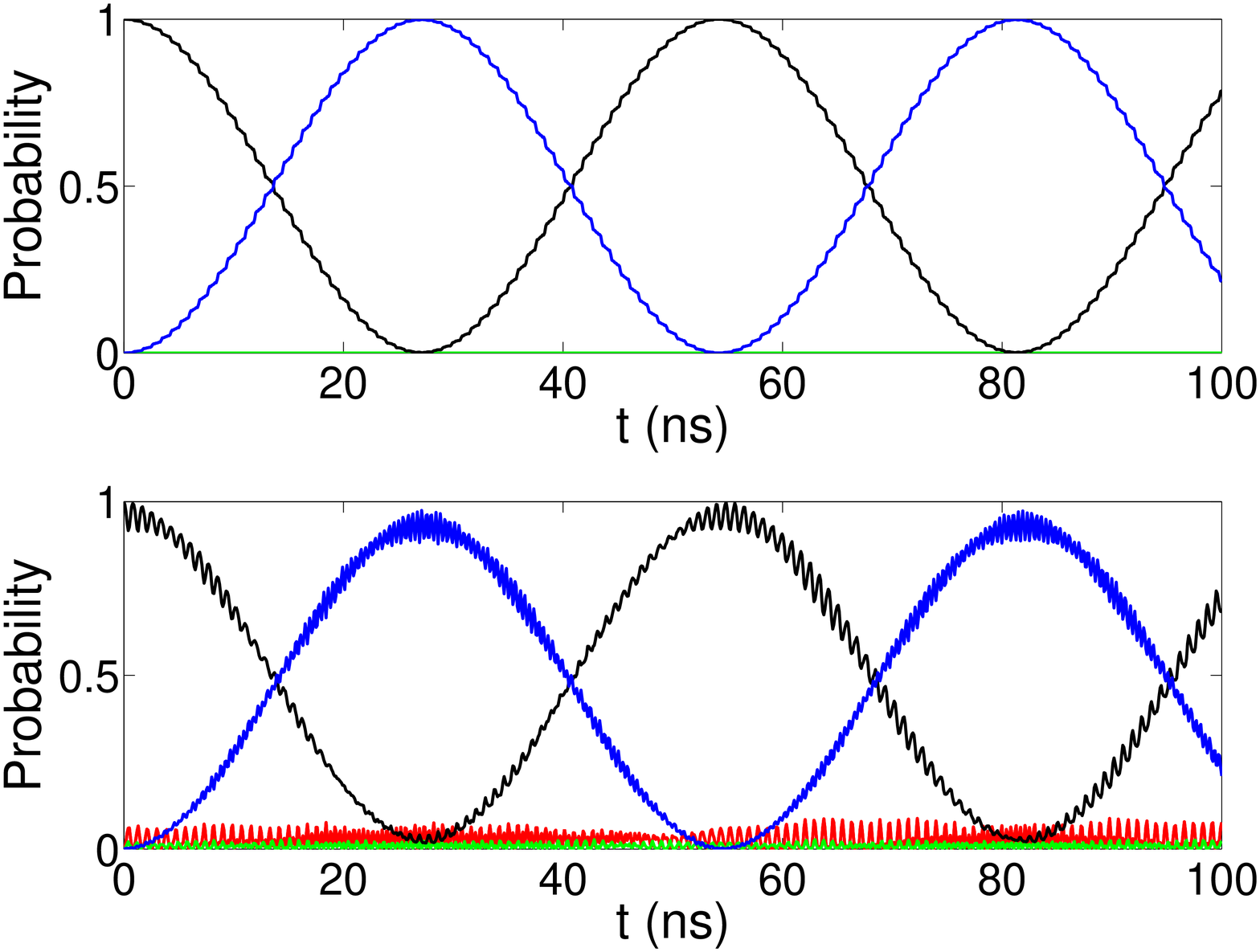}}
\end{picture}
\caption{\label{fg:sim1}
(Color online) Example of the coherent oscillations induced by a microwave at the frequency $(\tilde{\Delta}_2-\tilde{\Delta}_1)/h$. 
Above the two-qubit approximation was used whereas the lower plot is the result of the full three-qubit simulation.
The black line starting from probability 1 at $t=0$ corresponds to $|1_1 0_2\ket$
where as the blue line starting from probability 0 at $t=0$  corresponds to $|0_1 1_2\ket$.
The green and red lines correspond to the states  $|1_1 1_2\ket$ and $|0_1 0_2\ket$, respectively.
%The yellow line near unity is ${\rm Tr}\,\rho_{1,2}^2$. 
Here $\Delta t\approx 27.5$ ns results in gate equivalent to DCNOT. 
We use $\epsilon_3/h=7.375 $GHz. Here the microwave amplitude was $\delta\epsilon_3=$ 497 MHz which corresponds to 
Fig.~\ref{fg:stab}. With these choices $g_0\approx 0$ but $dg_\pm/d\Phi_3\neq 0$.
}
\end{figure}

As an illustrative example, let us consider the initial state $|1_1 0_2\ket$. 
Figure~\ref{fg:sim1} illustrates the probability of different outcomes in the single-qubit eigenbasis 
vs. time with the example parameters of Fig.~\ref{fg:J12}. \footnote{Should we plot the probabilities of the four lowest levels 
then the oscillation amplitude would be practically one. 
In the case of the effective Hamiltonian the used basis is of course slightly rotated from the original
and therefore the oscillation is perfect.}
In this example the DC component of the coupling is
$g_0\approx 0$. In this simulation we assumed that the microwave couples only to qubit 3.
However, even if the microwave crosscouples to all the qubit loops, it is still
possible to carry out the operations with high fidelity.
We see that after a time $h/(2g_-)$ the probabilities of the states $|1_1 0_2\ket$ and $|0_1 1_2\ket$ are flipped.
This corresponds to a gate locally invariant with the DCNOT, i.e. the operation in Eq.~(\ref{eq:DCNOT}).
The fact that the full three-qubit simulation shows reduced amplitude after tracing over the third qubit
is attributable to the fact that the third qubit entangles slightly with the primary qubits due to its finite frequency.
At best the fidelity ${F=\sqrt{\bra \psi|\rho_{1,2}|\psi\ket}}$ between the reduced density matrix and the target pure state 
${|\psi\ket=\exp(i\pi/4(\sigma_x^1\sigma_x^2+\sigma_y^1\sigma_y^2))|1_1 0_2\ket}$ is about 98\%. 

\begin{figure}
\begin{picture}(130,200)
\put(-75,0){\includegraphics[width=0.55\textwidth]{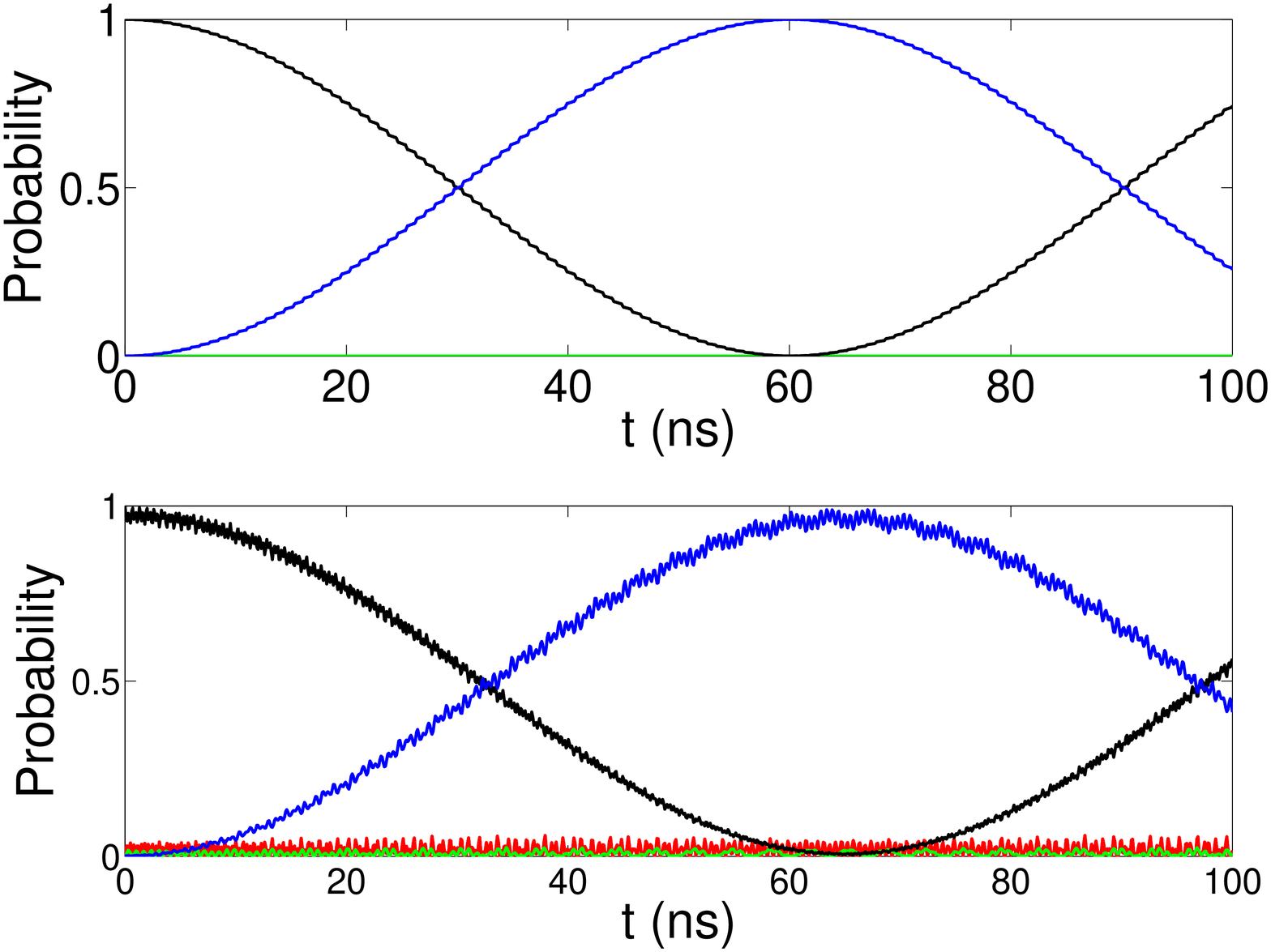}}
\end{picture}
\caption{\label{fg:sim2}(Color online) Same as in Fig.~\ref{fg:sim1} but with  $\Delta_3=15$ GHz and $\epsilon_3/h=5.73 $GHz.
The fidelity is at best above 99\%.
}
\end{figure}

The above example is not perfect but indicates clearly
that the scheme should be comparably easy to demonstrate in an experiment. To see how increasing $\Delta_3$ affects 
the situation, we consider the case  $\Delta_3=15$ GHz in Fig.~\ref{fg:sim2}. It can be seen clearly that while the gate time
stays reasonable (about 60 ns), the amplitude of the oscillation and thus the fidelity is improved above 99\%.

\begin{figure}
\begin{picture}(130,200)
\put(-75,0){\includegraphics[width=0.55\textwidth]{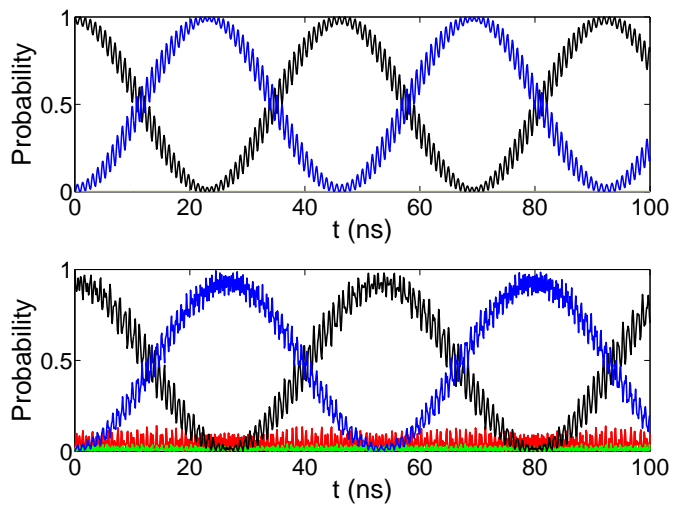}}
\end{picture}
\caption{\label{fg:sim3} (Color online) Evolution of the probabilities at the ``sweet spot'' of the coupling $dg_\pm/d\Phi_3=0$.
The parameters are like in Fig.~\ref{fg:sim1} otherwise, but 
$\epsilon_3/h=4.7$ GHz, $\epsilon_1=\epsilon_2=1.23$ GHz and $g_0/h=110$ MHz. Thus
again $\tilde{\epsilon}_1=\tilde{\epsilon}_2=0$. The fidelity is at best above 99\% and the gate time is about 26 ns.
}
\end{figure}

So far we have given examples of operation at $g_0\approx 0$. It is however possible to
work with finite $g_0$. Particularly, we would like to operate at $dg_\pm/d\Phi_3=0$. 
To demonstrate, we simulate the first example again at this point.
The results are shown in Fig.~\ref{fg:sim3}.
We now work at the frequency $\tilde{\omega}_-=\omega_- +2g_0^2/\omega_-$.
If this correction\cite{bertet} is not taken into account, the amplitude of oscillation is roughly halved
but with the corrected operation frequency, the best fidelity is again about 99\%. This is extremely promising since 
at this point the decoherence is expected to be reduced.

% see probs3_nocorr...

Since we are interested in creating a unitary gate, it is not sufficient to consider one initial state alone.
One possible measure is to calculate the fidelities for a complete set of initial states. It however turns out that the fidelities with 
the other logical qubit states for the present scheme are equally good as for the state $|1_1 0_2\ket$. This means that we can indeed carry out the
DCNOT with high fidelity.

We conclude that the present scheme is an experimentally realistic scheme for realizing two qubit gates at the optimal points
of the qubits.
We can also conclude that the derived two-qubit Hamiltonian is in good agreement with the full three-qubit Hamiltonian.
The desired behavior of the three-qubit system is only observed when the renormalization of $\epsilon$ and $\Delta$ 
are taken into account. In the case of a finite $g_0$ the operation frequency is further renormalized as can be seen in the example above.
The predicted gate time also agrees reasonably well with the full dynamics.

\section{Discussion}\label{sec:disc}

We have shown how to realize a tunable coupling scheme for optimally biased flux qubits using an extra qubit as a coupling element.
Although the mutual inductances are fixed, the primary qubits can be effectively coupled and decoupled at will
using microwaves at a suitable frequency.
As a first test of the scheme, 
we suggest the coupling to be demonstrated with just one measurement SQUID with all the three qubits sitting inside the loop.
The seeming visibility problem between $|0_11_2\ket$ and $|1_10_2\ket$ in this kind of a measurement can be overcome easily.
To realize coherent two-qubit oscillations of the kind presented in the previous Section and to see a signature of the tunable coupling,
one can convert the oscillation in the subspace spanned by $|0_11_2\ket$ and $|1_10_2\ket$
into an oscillation of $|0_10_2\ket$ and $|1_11_2\ket$ by just one $\pi$-pulse.
Scanning the MW frequency around $\omega_-$ should give adequate proof of the tunability.
This experiment could be carried out even with a continuous wave at $\omega_-$ and two chopped $\pi$-pulses, one to prepare either 
$|0_11_2\ket$ or $|1_10_2\ket$ and one to overcome the visibility problem. Scanning the interval between the pulses should result
in an oscillation between $|0_10_2\ket$ and $|1_11_2\ket$. 

For  more sophisticated control of the qubits, all the microwaves need to be chopped and attention needs to be paid to phases.
It is of course important that the signals are phase-coherent.
The choices made in this paper should be realizable in an experiment by properly adjusting the phase of the microwave source addressing the
qubits. Once this is done the exact timing of the coupling pulse is not important as time does not appear
explicitly in the rotating frame Hamiltonian. In the case of the FLICFORQ scheme timing is more crucial since
the rotating frame Hamiltonian still has an explicit time dependence.
The phases in Eq.~(\ref{eq:Hrot}) could of course be chosen differently from what we have done.
Like in FLICFORQ, it should be possible to utilize frequency multiplexing in the present coupling scheme.

Let us comment on the scalability of the present scheme. A generalized version of the presented transformation may be used to derive
an effective Hamiltonian for a larger multiqubit register without changing the form of the result.
For instance in the case of a linear chain of qubits with every second qubit acting as a coupling element the bias $\tilde{\epsilon}_j$
of any computational qubit will have a contribution from both of its neighboring
coupling elements and the same holds for $\tilde{\Delta}_j$. The coupling between neighboring computational qubits will be exactly the same
as in the text up to second order. We can therefore conclude that the scheme is potentially scalable to multiple qubits
since figuring out the controls required for a more complicated multiqubit unitary gate is a tractable task.

\acknowledgments
We thank P. Bertet for sending Ref.~[\onlinecite{bertet}] to us prior to publication.

\end{document}